\documentclass[11pt]{article}
\usepackage{times}
\usepackage{sidecap}
\usepackage{geometry}
\usepackage{floatrow}
\usepackage{float}
\usepackage{hyperref}
\usepackage{graphics,graphicx}
\usepackage{subcaption}
\usepackage{epsf,wrapfig}
\usepackage{natbib,bm}
\usepackage[misc]{ifsym}

\usepackage{paralist}
\let\olditem\item
\renewenvironment{thebibliography}[1]{%
  \section*{\refname}
  \let\par\relax
  \renewcommand{\item}[1][]{\olditem[\textbullet]}%
  \inparaenum}{\endinparaenum}

\newcounter{aff}

\DeclareGraphicsExtensions{.jpg,.pdf,.png,.eps}

\usepackage{caption}
\usepackage[dvipsnames]{xcolor}
\usepackage{tcolorbox}
\bibpunct{(}{)}{;}{a}{}{,}
\usepackage{soul}
\usepackage[version=4]{mhchem}

\definecolor{darkcyan}{rgb}{0.0, 0.0, 0.7}

\definecolor{boxcolor}{RGB}{193, 229, 255}
\definecolor{figboxcolor}{RGB}{193, 229, 255}
\definecolor{boxcolorgray}{RGB}{227, 227, 227}
\definecolor{darkcyan}{rgb}{0.0, 0.0, 0.7}

\sethlcolor{boxcolorgray}

\def\gtrsim{\lower 2pt \hbox{$\, \buildrel {\scriptstyle >}\over
{\scriptstyle \sim}\,$}}
\def\lesssim{\lower 2pt \hbox{$\, \buildrel {\scriptstyle <}\over
{\scriptstyle \sim}\,$}}

\def\approxlt{\lower.2em\hbox{$\buildrel < \over \sim$}}
\def\approxgt{\lower.2em\hbox{$\buildrel > \over \sim$}}

\def \ms           {\hbox{M$_{\odot}$}}

\newcommand{\jpb}   {Jy~beam$^{-1}$}    

\newcommand{\kms}   {km~s$^{-1}$}

\newcommand{\mj}    {M$_\mathrm{Jup}$}
\newcommand{\mearth}    {M$_\oplus$}
\newcommand{\Mdisk}  {M$_\mathrm{disk}$}
\newcommand{\Mstar}  {M$_\mathrm{star}$}
\def\arcmin{$^{\prime}$}
\def\arcsec{$^{\prime\prime}$}

\def\msun{M$_{\odot}$}
\def\lsun{L$_{\odot}$}

\def\kms{km\,s$^{-1}$}
\def\be{\begin{equation}}
\def\ee{\end{equation}}
\def\bea{\begin{eqnarray*}}
\def\eea{\end{eqnarray*}}

\def\jref@jnl#1{{\rm#1}}
\def\nar{\jref@jnl{New Astron. Rev.}}			
\def\aj{\jref@jnl{AJ}}			
\def\araa{\jref@jnl{ARA\&A}}		
\def\apj{\jref@jnl{ApJ}}			
\def\apjl{\jref@jnl{ApJ}}		
\def\apjs{\jref@jnl{ApJS}}		
\def\ao{\jref@jnl{Appl.~Opt.}}		
\def\apss{\jref@jnl{Ap\&SS}}		
\def\aap{\jref@jnl{A\&A}}		
\def\aapr{\jref@jnl{A\&A~Rev.}}		
\def\aaps{\jref@jnl{A\&AS}}		
\def\azh{\jref@jnl{AZh}}			
\def\baas{\jref@jnl{BAAS}}		
\def\jrasc{\jref@jnl{JRASC}}		
\def\memras{\jref@jnl{MmRAS}}		
\def\mnras{\jref@jnl{MNRAS}}		
\def\pra{\jref@jnl{Phys.~Rev.~A}}	
\def\prb{\jref@jnl{Phys.~Rev.~B}}	
\def\prc{\jref@jnl{Phys.~Rev.~C}}	
\def\prd{\jref@jnl{Phys.~Rev.~D}}	
\def\pre{\jref@jnl{Phys.~Rev.~E}}	
\def\prl{\jref@jnl{Phys.~Rev.~Lett.}}	
\def\pasp{\jref@jnl{PASP}}		
\def\pasj{\jref@jnl{PASJ}}		
\def\qjras{\jref@jnl{QJRAS}}		
\def\skytel{\jref@jnl{S\&T}}		
\def\solphys{\jref@jnl{Sol.~Phys.}}	
\def\sovast{\jref@jnl{Soviet~Ast.}}	
\def\ssr{\jref@jnl{Space~Sci.~Rev.}}	
\def\zap{\jref@jnl{ZAp}}			
\def\nat{\jref@jnl{Nature}}		
\def\iaucirc{\jref@jnl{IAU~Circ.}}
\def\aplett{\jref@jnl{Astrophys.~Lett.}}
\def\apspr{\jref@jnl{Astrophys.~Space~Phys.~Res.}}
\def\bain{\jref@jnl{Bull.~Astron.~Inst.~Netherlands}}
\def\fcp{\jref@jnl{Fund.~Cosmic~Phys.}}
\def\gca{\jref@jnl{Geochim.~Cosmochim.~Acta}}
\def\grl{\jref@jnl{Geophys.~Res.~Lett.}}
\def\jcp{\jref@jnl{J.~Chem.~Phys.}}	
\def\jgr{\jref@jnl{J.~Geophys.~Res.}}	
\def\jqsrt{\jref@jnl{J.~Quant.~Spec.~Radiat.~Transf.}}
\def\memsai{\jref@jnl{Mem.~Soc.~Astron.~Italiana}}
\def\nphysa{\jref@jnl{Nucl.~Phys.~A}}
\def\physrep{\jref@jnl{Phys.~Rep.}}
\def\physscr{\jref@jnl{Phys.~Scr}}
\def\planss{\jref@jnl{Planet.~Space~Sci.}}
\def\procspie{\jref@jnl{Proc.~SPIE}}

\def\arcmin{\hbox{$^\prime$}}
\def\arcsec{\hbox{$^{\prime\prime}$}}

\def\ion#1#2{#1$\;${\small\rm\@Roman{#2}}\relax}

\newcount\lecurrentfam
\def\LaTeX{\lecurrentfam=\the\fam \leavevmode L\raise.42ex
\hbox{$\fam\lecurrentfam\scriptstyle\kern-.3em A$}\kern-.15em\TeX}
\def\sizrpt{
(\fontname\the\font): em=\the\fontdimen6\font, ex=\the\fontdimen5\font
\typeout{
(\fontname\the\font): em=\the\fontdimen6\font, ex=\the\fontdimen5\font
}}
\definecolor{dark}{rgb}{ 0.34, 0.09, 0.49}

\usepackage{fancyhdr,url}
\usepackage{transparent}
\usepackage{tablefootnote}
\usepackage{mathpazo}

\begin{document}

% Title and subtitle
\noindent
\begin{raggedright}
{\Huge\bfseries 
Bridging the gap between stars \& planets \\[0.5em]
\Large \textit{ The formation of brown dwarfs and free-floating worlds}
}
\end{raggedright}\\

%\vfill
% Author block
\noindent
Aina Palau$^{1*}$,
Dimitris Stamatellos$^{2}$, 
Matthew Bate$^{3}$, 
Manuel Zamora-Avil\'es$^{4}$, 
Gisela Ortiz-León$^{4}$,
O. Sergio Rojas-García$^{1}$,
Mariana G\'omez-Cuartas$^{1}$,
Nuria Huélamo$^{5}$, 
\'Alvaro Ribas$^{6}$,
Philippe Andr\'e$^{7}$,
Carlos Carrasco-González$^{1}$,
Eduardo Mart\'in$^{8}$,
Stella Tsilia$^{8}$,
Myriam Benisty$^{9}$,
Pamela Freeman$^{1}$,
Emmanuel Caux$^{10}$,
Audrey Coutens$^{10}$,
Ryohei Kawabe$^{11}$,
Pavel Kroupa$^{12}$,
Aleks Scholz$^{13}$,
Richard Parker$^{14}$,
Amelia Bayo$^{15}$,
Alejandro Santamaría-Miranda$^{16}$,
Swastik Chowbay$^{17}$,
Patricia Luppe$^{18}$,
Kengo Tomida$^{19}$,
José Juan González-Avilés$^{20}$,
\'Alvaro S\'anchez-Monge$^{21}$,
Luis Felipe Rodríguez$^{1}$,
Itziar de Gregorio-Monsalvo$^{22}$,
Stefano Facchini$^{23}$,
Ian Czekala$^{17}$, 
Fumitaka Nakamura$^{24}$,
Lukasz Tychoniec$^{25}$,
Genaro Su\'arez$^{26}$,
Ngoc Phan-Bao$^{27}$,
Adnan-Ali Ahmad$^{28}$,
Benoit Commerçon$^{28}$,
Ricardo Hern\'andez-Garnica$^{1}$,
Aurora Sicilia-Aguilar$^{29}$,
Claudia Toci$^{30}$,
Paola Pinilla$^{31}$,
Quentin Kral$^{32}$,
Yinuo Han$^{33}$,
Matthew De Furio$^{34}$,
William DeRocco$^{35}$,
Catherine Walsh$^{36}$,
Luca Matr\`a$^{18}$,
%Endorsers
Herv\'e Bouy$^{37}$,
Adam Burgasser$^{38}$,
Catherine Espaillat$^{39}$,
Daniel Harsono$^{40}$,
Masahiro N. Machida$^{41}$,
Koraljka Muzic$^{42}$,
Rakesh Pandey$^{43}$,
Ignasi Ribas$^{21}$,
Carlos Rom\'an-Z\'u\~niga$^{44}$,
Jeremy Smallwood$^{45}$, 
Anthony Whitworth$^{46}$,
Luis Zapata$^{1}$,
David Blank$^{47}$\\

\medskip
{
\footnotesize{
\noindent $^*$ a.palau@irya.unam.mx \\ 
\noindent Affiliations are listed at the end of the text
}
}

\medskip

\begin{tcolorbox}[boxsep=3pt,left=0pt,right=0pt,top=1pt,bottom=1pt,colframe=boxcolor,colback=boxcolor]
\justify
\textbf{Abstract:} 
%Limit of 1920 characters
The formation mechanisms of brown dwarfs (BDs) and free-floating planets (FFPs) are poorly constrained. The two main competing scenarios include the star-like scenario, where these objects correspond to the least massive gravitationally unstable fragments of molecular clouds, and the planet-like scenario, where these objects form in a circumstellar disk and are subsequently ejected after a dynamical interaction.
A new generation millimeter interferometer, such as ALMA2040, is required to disentangle the relevance of these scenarios. For this, ALMA2040 must provide the appropriate performance to explore BD/FFP formation from the cloud scale ($\sim10$~pc) down to the disk scale ($\sim10$~au). 
At cloud scales, ALMA2040 must detect a $\sim1$~\mj\ embedded FFP in the outer Galaxy ($\sim10$~kpc) in 1 h, or the same object in nearby clouds ($\lesssim400$~pc) in about tens of seconds so that large-scale surveys of entire clouds can be performed. This would provide key censuses to compare the results with the predictions from each scenario.
At disk scales, ALMA2040 must detect CO outflows from first hydrostatic cores in the substellar regime (predicted to exist in the star-like scenario) and potential molecular outflows from $\sim1$~\mj\ FFPs in nearby clouds (for which outflow momentum rates are predicted in the star-like case). 
In addition, the planet-like scenario predicts a disk mass-to-star mass ratio for BDs larger than the star-like case, and testing this requires detecting disks of $\sim2$~\mearth\ around young BDs. 
Furthermore, if BDs form like stars, the predicted emission of complex organic molecules in proto-BDs at 400~pc must be detected with ALMA2040, requiring a factor of 10 increase in line sensitivity at a spatial resolution of 5--10 au. 
Finally, ALMA2040 must detect Earth-like protoplanets around young BDs at 400 pc, in order to provide key observational constraints on the radial drift behaviour and  planet formation mechanisms in the substellar regime.

\end{tcolorbox}

\newpage{}

\vspace{-0.2cm}
\section{Scientific context and open questions}
%Here is an example citation to show the style: \cite{Adamo2025}.

Brown dwarfs (BDs, with masses in the range 13--75~\mj) and free-floating/rogue planets (FFPs, with masses $\leq13$~\mj) constitute clues to build a full and comprehensive picture of both star and planet formation, with the consequent implications on the origin of life and life emergence probability in the Galaxy. However, the formation pathways, star-like or planet-like, of these bridging objects are poorly understood. 
%Similarly, their late stages of evolution are poorly studied. 
In the star-like pathway, these objects result directly from the collapse and fragmentation of molecular clouds, while in the planet-like pathway, these objects are formed from accretion in a circumstellar disk \citep[e.g.,][and references therein]{Palau2024}.
Progress has been slow mainly because current interferometers lack the appropriate sensitivity: searches are limited to nearby clouds ($\lesssim400$ pc), and even there these faint objects are only marginally detected.
For this reason, BD and FFP formation is one of the astrophysical areas where a next generation millimeter-wave interferometer, e.g., ALMA2040, could have a very high impact. With the increase of one order of magnitude in sensitivity and x3 in angular resolution, ALMA2040 would allow us to extend our current exploration to distances up to  several kpc, approximately 10 times farther than those reached by current studies. In the following sections, we provide potential science cases that ALMA2040 will revolutionize.
%Complement w Sec. 9 of Palau+24?

\section{BD and FFP formation: the star-like case}

\begin{figure*}[ht!]
\begin{center}
\includegraphics[width=1\columnwidth]{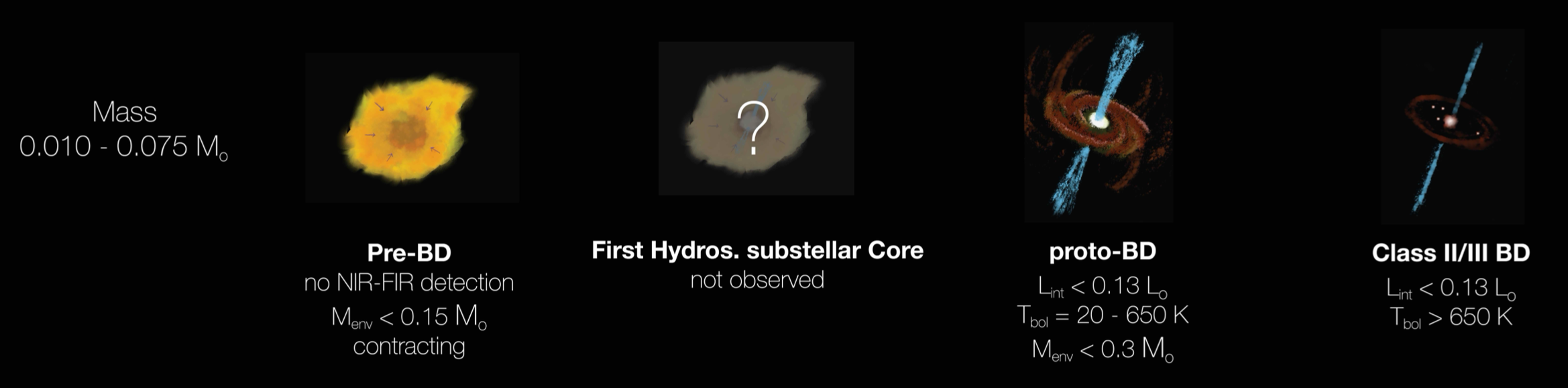}
\end{center}
\caption{Evolutionary phases of BD/FFP formation expected if these are formed in a star-like scenario, i. e., as a scaled-down version of low-mass stars \citep{Palau2024, Ahmad2025}.
}
\label{fig:Palau24-phases} 
\end{figure*}

One of the most promising scenarios for BD and FFP formation assumes that these very low-mass objects form like stars, that is, as a result of the fragmentation process of a molecular cloud \citep[MC, e.g.,][]{Bate2019, Ahmad2025}. 
%If the dominant formation mechanism of BDs and (high-mass) FFPs is the star-like scenario, two main predictions should be fulfilled: 
This model predicts that: 1) numbers of pre/proto-BDs/FFPs in MCs should be consistent with the results of simulations of star formation in clouds that fragment down to the opacity limit; and 2) BDs and FFPs should undergo the same phases of low-mass star formation scaled-down. These phases are summarized in Fig.~\ref{fig:Palau24-phases} and include: the pre-BD phase; the First Hydrostatic BD phase; the proto-BD phase (equivalent to the Class 0 and I phases of low-mass star formation); and the quasi-static contraction phase (including the Class II and III phases of low-mass star formation).
In the following, each one of these predictions are developed.

\subsection{Number of proto-BDs/FFPs in star-forming MC simulations}\label{sec:MCnumbers}

A powerful test for the star-like scenario is comparing the number of detected proto-BDs/FFPs in MCs with the expected number from MC simulations tracing fragmentation down to the opacity limit (typically of 3--10~\mj, \citealt{Boyd2005_opacitylimitedfrag}), such as those conducted by \cite{Bate2019}. In particular, the role of environment can be explored by comparing observations with predictions from numerical simulations. For example, it has been proposed that a slightly higher average cloud temperature, possibly raised by the feedback from nearby OB stars, could suppress the formation of objects in the low-mass end of the Initial Mass Function (IMF) and favour larger binary separations \citep[e.g.,][]{Bouy2026_Multiplicity-Taurus-UpSco}.

Currently, proto-BD detections in nearby clouds at 300--400 pc are limited to a luminosity threshold of 0.04~\lsun\ \citep{Palau2024}, which corresponds to the BD high-mass regime of $\gtrsim70$~\mj\ (according to the internal luminosity vs dynamical mass relation reported in this work). These luminosity and mass thresholds correspond to the sensitivity of the extensive and complete proto-BD search by \cite{Kim2016_VeLLOs}, who systematically explored 18 solar neighbourhood clouds ($\lesssim400$~pc) using single-dish telescopes. However, since the size of these objects is very small, of $\lesssim400$~au or $\lesssim1$~arcsec at 400 pc \citep[e.g.,][]{Lee2018_L328_ProtoBD_ALMA}, single-dish observations are problematic because of severe beam dilution. Blind searches are not feasible with current ALMA because mapping 18 clouds with the required sensitivity to detect these faint objects is too time-consuming.

%\vspace{-2pt}
\begin{figure*}[ht!]
\begin{center}
%\includegraphics[width=0.37\columnwidth]{brown-dwarfs/CIM1443_xy_hires10au_map_sinks-BDs.png}
%\includegraphics[width=0.54\columnwidth]{brown-dwarfs/histogram-masses_10auhighres.jpg}
%VERSION CIRCULATED AMONG CO-AUTHORS
%\includegraphics[width=0.4\columnwidth]{brown-dwarfs/CIM1443xy_sinksadd_1kpc_interm-ext-uvrange50.png}
%\includegraphics[width=0.55\columnwidth]{brown-dwarfs/CIM1443xy_sinksadd_1kpc_interm-ext-uvrange50_zoom.png}
\includegraphics[width=0.41\columnwidth]{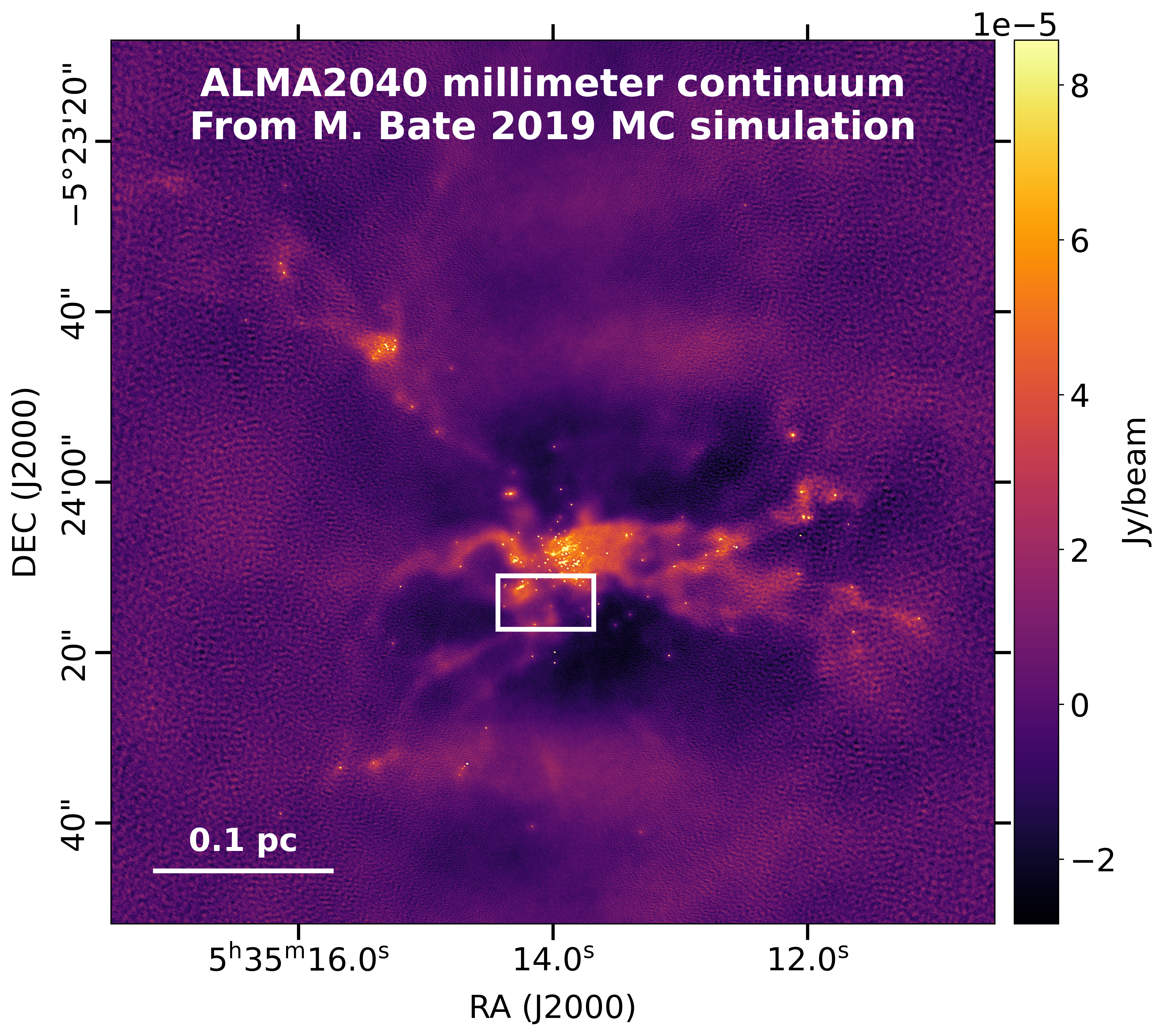}
\includegraphics[width=0.57\columnwidth]{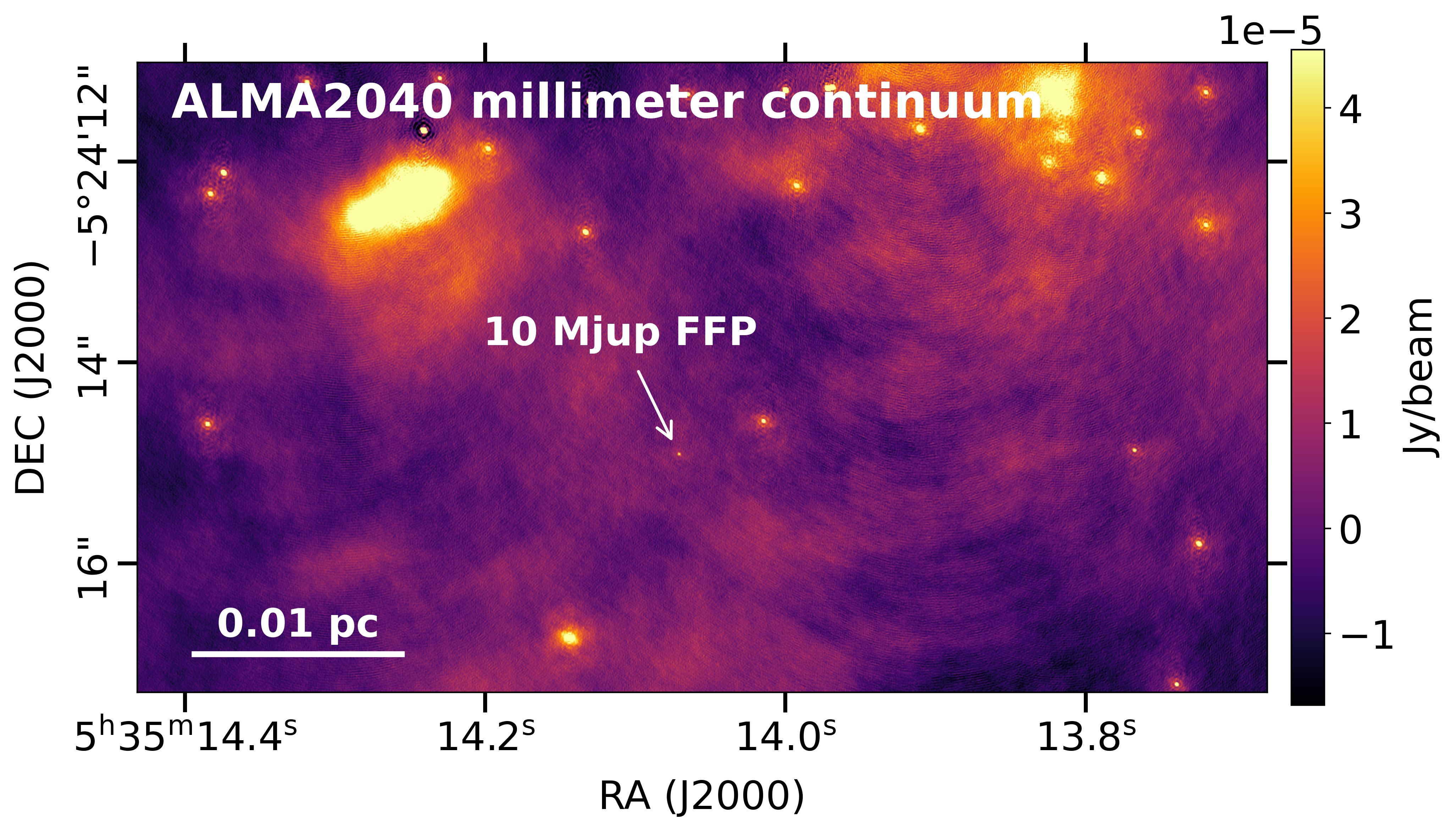}
\end{center}
%\vspace{-12pt}
\caption{ 
%{\bf Top-left:} H$_2$ column density from \cite{Bate2019} of a collapsing solar metallicity cloud, for the last snapshot at 1.3 times the free-fall time. Cyan crosses indicate BDs and white empty dots indicate stars.
%{\bf Top-right:} IMF from these simulations.
{\bf Left:} Synthetic observations at 3 mm of the simulated cloud at 1 kpc, for 1 h integration time and
a 255 12m antenna array with 512 GHz bandwidth and a synthesized beam of $\sim20$~mas.
%at Band 3 for intermediate (4h) + extended (4h) configurations combined and using 128 GHz of bandwidth. 
The synthetic observations are obtained from simulations of a collapsing solar metallicity cloud at the last snapshot corresponding to 1.3 times the free-fall time \citep{Bate2019}. 
%Cyan crosses indicate BDs and white empty dots indicate stars.
{\bf Right:} Zoom-in on one portion of the cloud marked with a white square on the left panel, where a FFP of 10~\mj\ is detected at 40~$\mu$\jpb, corresponding to a 20$\sigma$ detection. 
%Zoom: using only extended config 4h, 128 GHz -> equivalent to 1h 512 GHz: Restoring beam = 0.0293797" X 0.0260054", 45.0436 deg.
}
\label{fig:Bate} 
%\vspace{-12pt}
\end{figure*}

%Using simulations from BDs forming in entire molecular clouds, such as those presented in \cite{Bate2019}, we could estimate how many BDs would be expected to be identified in molecular clouds at a certain distance. 
In order to explore the proto-BD population that ALMA2040 could detect, we used the snapshot at 1.3 times the free-fall time from the \cite{Bate2019} simulations of a collapsing solar metallicity cloud and simulated an ALMA2040 observation. This allowed us to obtain synthetic observations of a nascent cluster down to the substellar regime. The cloud has 500~\msun\ with initial density of $6\times10^4$~cm$^{-3}$, and produces 66 BDs, 4 FFPs, and 188 stars, with a BD+FFP fraction of 27\%. The least massive object formed in this simulation has a mass of 6~\mj.
%The column density of the final snapshot is presented in Fig.~\ref{fig:Bate}-top left, and the resulting IMF is shown in Fig.~\ref{fig:Bate}-top right. 
%The 4 FFPs in the simulations have the following masses: 2 FFPs of 6 Mjup and 2 FFPs of 13 Mjup.
%In order to perform synthetic observations, the sinks were added to the column density map without sinks, with their corresponding mass converted to column density. 
To convert the column density of the simulation to intensity, we assumed optically thin thermal dust emission for a 20 K isothermal cloud, a gas-to-dust mass ratio of 100, and a dust opacity per unit mass of dust+gas at 3 mm of 0.003 cm$^2$\,g$^{-1}$ \citep{Birnstiel2018_dustopacity}.
The ALMA2040 synthetic observations at 3~mm (selected to have the appropriate field of view) are produced assuming a distance of 1 kpc, 1 hour of integration time and a hypothetical 255 12m antenna array with 512 GHz bandwidth and a synthesized beam of $\sim20$~mas.
%using the extended (4 h) {\bf antenna} configurations in Band 3. 
The resulting synthetic observation, presented in Fig.~\ref{fig:Bate}, recovers all the 70 BDs+FFPs formed in this simulation. 
%It is important to note that the cloud has been placed at 1 kpc, about an order of magnitude farther out than the typical nearby clouds where the proto-BD content has been studied so far \citep[e.g.,][]{Kim2016_VeLLOs, Palau2024}. 
In this case, the measured rms is 2~$\mu$\jpb, larger than the nominal rms for this setup (of 0.14~$\mu$\jpb) because of the extended emission of the cloud. 
The 5$\sigma$ mass sensitivity for the achieved rms corresponds to 3~\mj, well below the mass of the least massive object formed in the simulation. Assuming that the extended emission of the cloud is filtered using an appropriate antenna configuration, that we are observing at 230 GHz and that the nominal noise is reached for 1 h and 512 GHz of bandwidth (rms=0.2~$\mu$\jpb), the 5$\sigma$ mass sensitivity at 1 kpc is $\sim0.01$~\mj, or $\sim1$~\mj\ at 10 kpc. Thus, the dusty envelopes/disks of proto-BDs/FFPs could be detected out to the outer Galaxy with ALMA2040 (Table~\ref{tab:requirements}).

%MZA:
%En realidad todo el gas es gas denso, pues la densidad inicial es de 6e4 cm^-3.
%M_gas_total           = 500 Msun
%M_gas (sub cubo) = 181.3 Msun (todo el cubo: 373.3 Msun)
%M_sinks                   = 120.9 M☉ (256 sinks)  
%SFE                           = 40% 
%La sub-box en la que estamos trabajando es de 0.5 pc por lado.

In addition to the detection of proto-BDs/FFPs in clouds at kpc distances, ALMA2040 would also allow to perform deep searches in entire clouds of the solar neighbourhood down to $\sim1$~\mj. For this, large mosaics would be required and low frequencies are best suited to have larger primary beams. Assuming a distance of 400 pc, each pointing should integrate for 5 s to reach a 5$\sigma$ mass sensitivity of $\sim1$~\mj\ at 100 GHz (Table~\ref{tab:requirements}). For a cloud like Perseus, with an area of $\sim100$~pc$^2$,
%or $\sim3$~pc per side, 
171 pointings per side are needed at 400 pc (at almost Nyquist sampling with 30 arcsec of separation between pointings),
%the primary beam is around 52 arcsec), 
which would make 29241 pointings in total, requiring an observation time of $\sim40$~h. For a sample of 10 similar nearby clouds, a total of about 400 h would be needed, making an excellent case for an ALMA2040 Large Program. With ALMA in its current performance, this would require $\sim160000$ hours or 18 years of uninterrupted observations.

%CHECK IF YOU EXPECT A Npbd vs Mcloud relation as in \cite{Palau2024}.

\subsection{Scaled-down version of the known phases of star formation}

In this subsection, we assume that BDs/FFPs form like stars and that they must undergo the phases outlined in Fig.~\ref{fig:Palau24-phases} \citep[e.g.,][]{Ahmad2025}. This implies that pre-BD/FFP cores, first hydrostatic substellar cores, proto-BDs/FFPs and ClassII/III BDs/FFPs should exist, with protostellar properties scaled-down. We present below the requirements for ALMA2040 to detect the thermal dust and molecular emission of BDs/FFPs in each evolutionary phase. 
%The main phases of low-mass star formation are summarized in Fig.~\ref{fig:Palau24-phases} and include: 1) the pre-BD phase; 2) the First Hydrostatic BD phase; 3) the proto-BD phase (equivalent to the Class 0 and I phases of low-mass star formation); and 4) the quasi-static contraction phase (including the Class II and III phases of low-mass star formation).

\subsubsection{Pre-BDs/FFPs: pre-substellar cores}

The first step in the star formation process is known as the pre-stellar phase. In this phase, cores are already undergoing collapse but no hydrostatic core is formed yet. 
For the 500~\msun\ simulated cloud presented in Sec.~\ref{sec:MCnumbers}, about 70 BDs+FFPs are formed as a result of cloud fragmentation, implying that we would expect a similar number of pre-BDs/FFPs if they were formed simultaneously and the cloud was observed at the appropriate time. 
However, these cores have remained elusive, with very few candidates identified so far \citep[e.g.,][]{Andre2012_PreBD}. One of the most robust candidates, located in Ophiuchus and called Oph\,B-11, was detected at 3 mm using the IRAM PdBI after a single-dish search \citep{Andre2012_PreBD}. Future single-dish searches have been inefficient to find gravitationally bound very low-mass cores, probably because of beam dilution \citep[e.g.,][]{deGregorio2016_PreProtoBD_ChaII, SantamariaMiranda2021_ALMA_Lupus_ProtoBD}. Thus, it is not clear yet if the current lack of pre-BDs is due to an observational effect or because the main formation pathway is not star-like: if BDs/FFPs are formed in circumstellar disks, very few pre-BDs should be found. As shown in Sec.~\ref{sec:MCnumbers}, the high sensitivity of ALMA2040 would allow us to map entire MCs of the solar neighbourhood with a mass sensitivity of $\sim1$~\mj, a factor of 20 lower than the mass of Oph\,B-11. Thus, the identification of a significant number ($\sim100$) of pre-BDs/FFPs in blind searches in MCs would be consistent with the expectation for the star-like case (although detailed comparisons with simulations should be performed). Alternatively, if very few pre-BDs/FFPs are detected in a cloud, this would favour the planet-like scenario. 

It is important to remark that such faint pre-BDs/FFPs, if detected only in the millimeter continuum, could be confused with extragalactic objects. Ideally, molecular line detections at the position of the continuum source and velocities of the parental cloud would be required to confirm their nature. This further calls for a new-generation interferometer able to detect very faint line emission, which could be a factor of 20 fainter than the detections of current interferometers.  

\subsubsection{First Hydrostatic BD/FFP cores (FH-BDs)}

In this phase, the collapsing core reaches the density to transition from isothermal to adiabatic collapse. Because of the large opacities, the energy released during the collapse can no longer be radiated and is converted into heat, halting the collapse, and forming the first hydrostatic core (FHC). 
%There are several examples in the literature of these kind of objects \citep[e.g.,][]{Maureira2020_FHCs, Hirano2024A_FHC}. 
However, the corresponding stage for the substellar case has not been found (Fig.~\ref{fig:Palau24-phases}). 
%This is illustrated in the second stage of Fig.~\ref{fig:Palau24-phases}, where a question mark indicates that no nascent BD has been found at this stage yet \citep[e.g.,][]{Palau2024}.

The crucial property that allows to distinguish FHCs from pre-stellar cores is the fact that FHCs are already driving outflows 
%because the hydrostatic object is already formed with a pseudo-disk 
\citep[e.g.,][]{Machida2009_BDformation, Commercon2012_FHCs-SEDs}. In addition, because of their extremely early stage, the bolometric luminosity and temperatures are very low \citep[e.g.,][]{Commercon2012_FHCs-SEDs}. Based on the FHC model of \cite{Bhandare2018_FHC} for different initial cloud masses, the first-order properties expected for FH-BDs can be estimated. Extrapolating the mass outflow rates reported in Table 2 of that work from a 0.5 to 0.005~\msun\ core (as a fiducial value representative of a core forming a FFP), 
%for which a mass outflow rate of $3.5\times10^{-5}$~\msun\,yr$^{-1}$ is predicted.
a mass outflow rate of $10^{-8}$~\msun\,yr$^{-1}$ is estimated for FH-BDs.
%which should be constant until the proto-BD is formed and then decline \citep{Machida2009_BDformation}. 
This is about an order of magnitude lower than the mass outflow rate estimated in some (high-mass) proto-BDs \citep[e.g.,][]{Hsieh2016_Binary_ProtoBD_Jet_IRAS16253} and can be detected by ALMA2040 as shown in the next section. 
%As shown by \cite{Machida2009_BDformation}, the mass accretion rate stays constant during the FH-BD formation and the first 10$^4$~yr of the proto-BD. Specific simulations of FH-BD would allow to provide tighter constraints on the expected properties of FH-BDs.
%Table~4 of \cite{Palau2014} provides a summary of some of the FHC properties, being a very low bolometric temperature one of their most relevant properties, along with very low velocity outflows (and, consequently, very low outflow momentum rates).

\subsubsection{Proto-BDs/FFPs: embedded substellar objects in the Class 0/I phase}

\begin{figure*}[ht!]
\begin{center}
\includegraphics[width=1.\columnwidth]{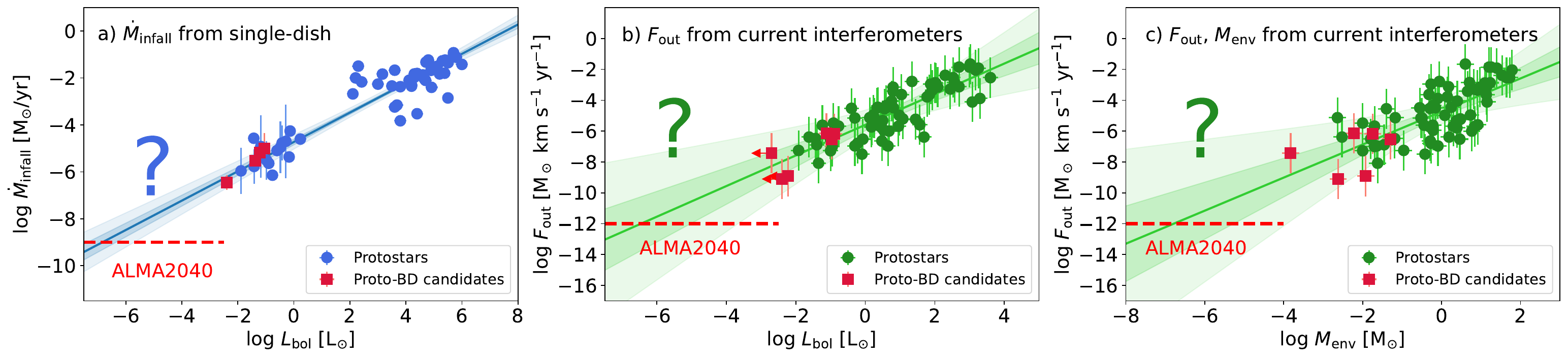}
\end{center}
\caption{Well-known correlations (mass infall rate vs bolometric luminosity (a), outflow momentum rate vs bolometric luminosity (b) and vs envelope mass (c)) for protostellar objects relating their accretion/ejection properties. If BDs are formed as scaled-down version of low-mass protostars, they should follow the low-end of the extrapolation \citep{Palau2024}. The dashed red horizontal line indicates the ALMA2040 detection limit.
}
\label{fig:Palau24-ysocorrels} 
\end{figure*}

Once the FHC (protostar) is formed, the main accretion phase and its subsequent immediate phase are referred to as Class 0 and I phases. At these stages, correlations between the accretion/ejection properties and the envelope mass or bolometric luminosity are well-established for protostars. Three of these correlations are shown in Fig.~\ref{fig:Palau24-ysocorrels}. If BDs/FFPs form as scaled-down versions of low-mass stars, the (Class 0/I) proto-BDs/FFPs should lie at the low-end extrapolation of these correlations. However, the current ALMA sensitivity prevents us from fully exploring this regime, as shown by the red-square symbols of the figure, corresponding to the current interferometric proto-BD detections \citep[e.g.,][]{Palau2024}. Fig.~\ref{fig:Palau24-ysocorrels} shows that the currently detected proto-BDs are of high mass and present significant dispersion possibly due to evolutionary effects, environmental conditions, opacity and inclination uncertainties. Thus, a new generation interferometer is required that fully covers the low-mass end of the BD regime and the FFP regime.

Ideally, ALMA2040 should be able to detect molecular outflows driven by proto-FFPs of $\sim1$~\mj\ and $\sim10^{-5}$~\lsun. This corresponds to mass outflow rates of $5\times10^{-11}$~\msun\,\kms\,yr$^{-1}$ assuming that the known protostellar relations of Fig.~\ref{fig:Palau24-ysocorrels} can be extended to lower masses. Assuming a weak and compact CO\,(2--1) outflow of $\sim200$ au, with a lobe 0.5~\kms\ wide, this outflow momentum rate corresponds to a line peak of of 1.5~m\jpb, for which an rms$\sim0.3$~m\jpb\ is required. This can be achieved with ALMA2040 for 1 h integration time and channel width of 0.26~\kms\ (Table~\ref{tab:requirements}). If longer integration times are used and somewhat different parameters are considered, ALMA2040 could detect outflow momentum rates as low as $5\times10^{-12}$~\msun\,\kms\,yr$^{-1}$.
%(of $\sim1$~'') 
%with CO\,(2--1) intensity of 15 m\jpb, rms of 3 m\jpb\ is required to detect it, and this can be achieved with current ALMA for 1 h integration time for a channel width of 0.26~\kms. 
%0.2MHz -> 0.26 km/s
%Assuming that the outflow lobe is 0.5~\kms\ wide,
%%as in the case of J401757B
%this corresponds to $5\times10^{-10}$~\msun\,\kms\,yr$^{-1}$, of the order of the lowest outflow momentum rate measured for proto-BDs (red symbols) of panels b and c of Fig.~\ref{fig:Palau24-ysocorrels}. 
%For 1 h integration time and channel width of 0.26~\kms with ALMA2040, outflows with CO(2--1) emission of 1.5~m\jpb\ could be detected (rms$\sim0.3$~m\jpb), corresponding to mass outflow rates of $5\times10^{-11}$~\msun\,\kms\,yr$^{-1}$. 
%If BDs/FFPs follow the known correlation between outflow momentum rate and bolometric luminosity, ALMA2040 would detect outflows driven by proto-FFPs of $2\times10^{-5}$~\lsun, corresponding to a dynamical mass of $\sim1$~\mj.
%log10(5e-11) = -10.301029995664
%log10(5e-11)+5.59)/0.99 = -4.7586161572363 -> Palau+24 Pout vs Lbol relation
%10**(-4.76) = 1.7378008287494E-05 Lsun
%
Therefore,
%With the improvement in sensitivity by more than one order of magnitude,
ALMA2040 will provide enough statistics to populate the extreme low-luminosities, required to properly test if the lowest luminosity objects follow the known relations for protostars.

%Finally, a common feature of protostellar objects revealed by ALMA is the presence of accretion streamers \citep[e.g.,][]{Pineda2020_streamer}.   
%\cite{Riaz2024_streamer} report an accretion streamer in HCO+(3--2) of integrated flux of 32 mJy towards a proto-BD of 0.04~\lsun, showing that proto-BDs can share this feature with protostars. With a rms in Band 6 of 0.3 m\jpb\ at 0.26~\kms\ for 1 h integration time, ALMA2040 would detect accretion streamers about one order of magnitude fainter, constituting an important test for the star-like scenario.

\subsubsection{Class II/III BDs/FFPs}

In the Class II/III phase, young stellar objects have negligible envelopes and are characterized by a well defined disk+outflow structure. A number of studies have been conducted to characterize the disks of Class II/III BDs. These studies consistently find disk masses in the range 0.02--3~\mj, and \Mdisk/\Mstar\ ratios around 0.001 \citep[e.g.,][]{Testi2016_BDdisks, Rilinger2019_BD-disks, Rilinger2021_BD-disks, AlmendrosAbad2024_Maccrate-Mstar-substellar}, although this must be an upper limit if non-detections are taken into account.
%\cite{Testi2016_BDdisks}: 17 BDs disks in Oph, masses from 50--600~Earth masses of gas+dust (assuming a gas-to-dust mass ratio of 100), corresponding to 0.16--1.9~\mj.
%\cite{Rilinger2019_BD-disks} model:  0.2--2~\mj, consistent with formation through a condensation from a (large) mass reservoir
%\cite{Rilinger2021_BD-disks}: several nearby clouds find total disk masses 0.02--3~\mj
% \cite{AlmendrosAbad2024_Maccrate-Mstar-substellar}: 26 Class II BDs with masses in the range 
Integrating for 1 h with ALMA2040, at 1~mm and 512 GHz of bandwidth, a rms of 0.2~$\mu$\jpb\ is achieved. This corresponds to a minimum 5$\sigma$ detectable mass of 2~\mearth\ at 1 kpc (assuming a dust temperature of 20 K and a dust opacity of dust+gas of 0.03~cm$^2$\,g$^{-1}$, \citealt{Birnstiel2018_dustopacity}). Thus, the same type of studies that have been conducted with ALMA in the solar neighbourhood will become feasible up to 1 kpc with ALMA2040. The improvement in BD disk statistics, including a large variety of environments, will allow us to tightly constrain the BD disk properties. 
Furthermore, recent works have also reported the presence of disks around FFPs of masses $\sim10$~\mj\ \citep[e.g.,][]{Bayo2017_IFFP, Scholz2023_NGC1333}. Assuming \Mdisk/\Mstar\ $\sim0.01$ for a proto-Neptune of $\sim0.05$~\mj, the associated disk could be detected up to 200 pc with ALMA2040.
% 0.05 Mjup * 0.01 = 5e-4 Mjup: 0.002*200*200/(400*400)=5E-4 Mjup , where 0.002 Mjup is the 5sigma mass sensitivity at 400pc in Band 6 and 128 GHz

%\begin{figure*}[ht!]
%\begin{center}
%\includegraphics[width=0.4\columnwidth]{brown-dwarfs/mdisk_mstar.pdf}
%\end{center}
%\caption{Updated $Mdisk$ vs $Mstar$ relation from Figure 3 of \cite{Stamatellos2015}, as published in Hern\'andez-Garnica et al. (2026). UPDATE WITH ALMENDROS-ABAD+24 FITS TO DIFFERENT CLOUDS.
%}
%\label{fig:MdiskMstar} 
%\end{figure*}

\subsubsection{Astrochemistry in young BDs/FFPs}

Astrochemistry in BDs is increasingly being explored from the proto-BD phase through the Class~II stage, revealing a chemically active environment despite the low luminosities of these objects.
Single-dish observations of proto-BDs show detections of multiple species (e.g.,  H$_2$CO, HCO$^+$, CS, CN, HNC, CH$_3$OH) and several deuterated molecules \citep[e.g.][]{Riaz2018_chem-OphSer, Riaz2019b_chem-OphSer, Riaz2022_HDCO_D2CO_Class0_I_ProtoBD,Riaz2022_D_H_ratio_ProtoBD, Riaz2023_Methanol}.
However, while the limited angular resolution and beam dilution of single-dishes makes it difficult to unambiguously locate the origin of the detected emission, interferometric studies of chemistry in proto-BDs are very scarce \citep[e.g.,][]{Anderl2020_Chemistry_VeLLO_IRAM04191}. %ALMA2040 interferometric observations, by providing superb angular resolution and sensitivity, are therefore crucial for disentangling the different emitting components and establishing the spatial distribution and physical origin of the reported molecular emission.
%Interferometric studies on very low-mass objects have allowed to study episodic accretion in the substellar domain, although these studies are very scarce \citep[e.g.,][]{Hsieh2018_VeLLOs, Anderl2020_Chemistry_VeLLO_IRAM04191, Kulterer2024_outburst}.
%specially relevant during the Class\,0 stage, with N$_2$H$^+$ and CO lines being potentially tracing shifts in the CO snowline related to accretion bursts \citep[e.g.,][]{Hsieh2018_VeLLOs, Anderl2020_Chemistry_VeLLO_IRAM04191}, or the NO molecule being a potential burst tracer \citep{Kulterer2024_outburst}. 
%, when substellar objects accrete most of their mass, ultimately shaping their evolutionary fate. 
%A similar strategy was followed by \citet{Anderl2020_Chemistry_VeLLO_IRAM04191} to conclude that the VeLLO IRAM04191+1522 probably underwent an accretion burst less than 1000 years ago. In this context, the NO molecule has been also proposed as a potential outburst tracer \citep{Kulterer2024_outburst}.
 
\begin{figure*}[ht!]
\begin{center}
\includegraphics[width=1.\columnwidth]{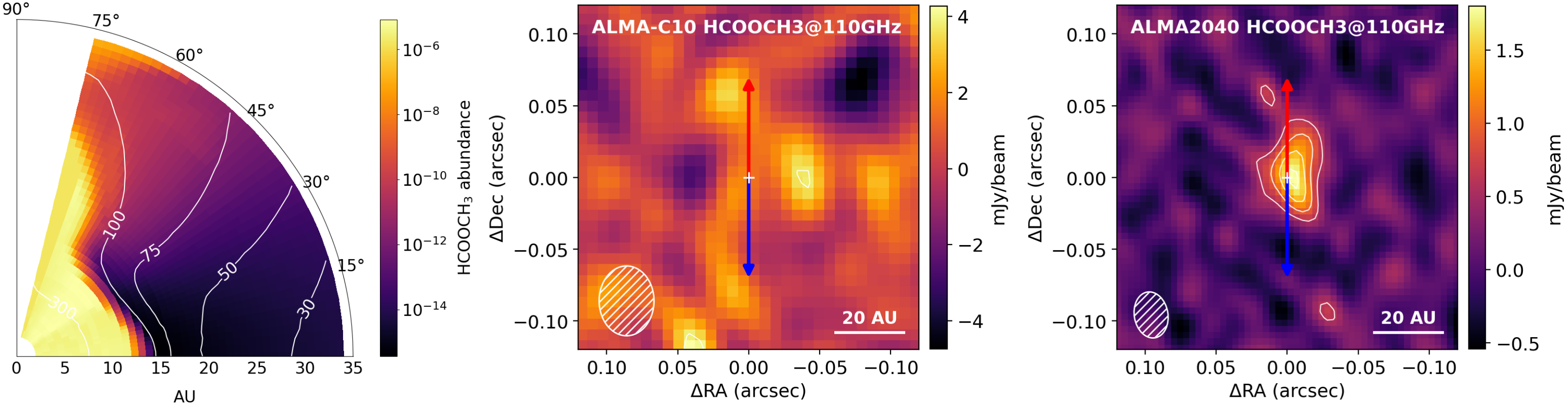}
\end{center}
%\vspace{-12pt}
\caption{
{\bf Left:} APE simulations of the expected abundance distribution of a 0.1~\msun\ core forming a 0.07~\msun\ Class 0 proto-BD for HCOOCH$_3$ at 110.8 GHz. 
{\bf Center + right:} Synthetic observations of 6 h integration for ALMA-C10 (center) and ALMA2040 (right) of the HCOOCH$_3$ transition at 110.8 GHz of the mid-left panel, for the 0.4~\kms\ width channel where the line peaks. While ALMA-C10 cannot detect the emission, ALMA2040 can resolve the emission elongated along the outflow direction and detects it at $\sim10\sigma$.
}
\label{fig:APE} 
%\vspace{-12pt}
\end{figure*}

%As it follows, all these studies give clues to understand the physical conditions and formation processes of proto-BDs.
% COMS: methanol
%Finally, the search for complex organic molecules (COMs), considered potential precursors to more complex prebiotic chemistry during planet formation, has also been extended to the substellar regime, with methanol being detected in one case \citep{Riaz2023_Methanol}.
%reported the single-dish detection of methanol (CH$_3$OH) in a sample of Class 0/I proto-BDs, providing evidence that COMs are already present during the earliest stages of substellar evolution.
%
To show the capability of ALMA2040 to explore chemistry in proto-BDs, we used the Astrophysical Protostellar Environment code \citep[APE,][]{Marchand2025_APE} to obtain an abundance map of the Complex Organic Molecule (COM) HCOOCH$_3$. The code assumes that a core of 0.1~\msun\ collapses in a star-like manner forming a 0.07~\msun\ proto-BD at 400 pc, and the resulting abundance map is presented in Fig.~\ref{fig:APE}-left. In this code, a gravitationally unstable core naturally forms a disk. The code includes cavity outflows that are heated by the radiation propagating through the cavity. The dynamical code is coupled to the chemical code {\it Nautilus} \citep{Ruaud2016_Nautilus} to provide abundance distribution maps. As can be seen in Fig.~\ref{fig:APE}, HCOOCH$_3$ is expected to be found distributed along the outflow cavity walls and in the disk, with abundances around 10$^{-6}$ up to 20 au in the vertical direction. 
%Assuming an envelope H2 column density of $2\times10^{24}$~cm$^{-2}$ (adopting 30~\mj\ within 200 au, e.g., \citealt{Palau2014}), this corresponds to N(CH3OCH3)$\sim2\times10^{-18}$ and N(HCOOCH3)$\sim2\times10^{-19}$. 
% say '0.03*1.9891e33/(2.8*1.6733e-24)/(pi*(100*1.4960e13)**2)' 
%1.8114776077966E+24
While the current ALMA interferometer cannot detect the HCOOCH$_3$ emission at 110.8 GHz in 6 h of integration and for a channel width of 0.4~\kms\ (Fig.~\ref{fig:APE}-center), ALMA2040 could detect it at 10$\sigma$ (rms$=0.2$~m\jpb), and resolve it elongated along the outflow direction (Fig.~\ref{fig:APE}-right), for the same integration time and channel width. 
%In 6 hours, the rms would be 50~$\mu$\jpb, allowing to detect the HCOOCH$_3$ emission from proto-BDs/FFPs at 400 pc. 
Thus, ALMA2040 will open for the first time the study of COMs and hot corinos associated with proto-BDs/FFPs. 

%ALMA2040 will allow to extend all these studies to larger samples, shedding light on their formation processes.

%\vspace{-2pt}

% Outburst
%Interestingly, chemistry has also been used as a tool to study episodic accretion in the substellar domain.  This is specially relevant during the Class\,0 stage.
%, when substellar objects accrete most of their mass, ultimately shaping their evolutionary fate. 
%As an example, \citet{Hsieh2018_VeLLOs} studied episodic accretion in a sample of Very Low Luminosity Objects (VeLLOs) analysing the spatial distribution of the line emission from N$_2$H$^+$ and CO isotopologues. In some objects, they show that the N$_2$H$^+$ emission was depleted toward the center of the sources, while the CO isotopologue emission exhibited a peak. The authors proposed that this anticorrelation is caused by the shift of the CO snowline due to accretion bursts, and showed that these episodes seem more frequent in VeLLOs that in low-mass protostars. A similar strategy was followed by \citet{Anderl2020_Chemistry_VeLLO_IRAM04191} to conclude that the VeLLO IRAM04191+1522 probably underwent an accretion burst less than 1000 years ago. In this context, the NO molecule has been also proposed as a potential outburst tracer \citep{Kulterer2024_outburst}.

\section{BD and FFP formation: the planet-like case}\label{sec:ejections}

%In the previous sections, specific science cases have been presented that would be crucial to test the star-like scenario of BD and FFP formation. In this section we briefly diskuss the planet-like+ejection scenario. 
This scenario considers that a substellar object formed in a circumstellar disk interacts with other siblings and is eventually ejected. While ejected FFPs could have formed by core accretion in a circumstellar disk, the more massive BDs are not easily formed through core accretion because it would require very long timescales, and one of the most plausible mechanisms is disk fragmentation \citep[e.g.,][]{Stamatellos2009}.
According to simulations and observational data, this scenario could be as relevant as the star-like scenario.
Considering the simulated cloud used in Sec.~\ref{sec:MCnumbers}, which produces 188 protostars, about 37 protostars (in the Class 0/I phase) should have disks massive enough (with \Mdisk/\Mstar\,$\gtrsim0.3$) to be gravitationally unstable, assuming that 20\% of the protostars have these massive disks \citep[e.g.,][]{Tobin2020_VANDAM-Orion, Nazari2025_planets-disks-substructures}.
%a fraction of stars with disks fulfilling the disk fragmentation criterion of Mdisk/Mstar $>0.3$ \citep[e.g.,][]{MercerStamatellos2020} 
According to \cite{Stamatellos2009}, the average number of substellar objects formed by disk fragmentation per star is in the range 5--10, implying that a total of 185--370 substellar objects should be formed by disk fragmentation in the cloud of Sec.~\ref{sec:MCnumbers}. Assuming an ejection fraction of 55\% \citep[e.g.,][]{Stamatellos2009}, the total number of ejected substellar objects in the cloud of Sec.~\ref{sec:MCnumbers} should be 100--200. 
%188*0.2*5*0.55=103.4
%188*0.2*10*0.55=206.8
This is of the order of the number of BDs formed via the star-like scenario (of 70 in that simulation). These approximate numbers show that the disk fragmentation scenario cannot be disregarded when considering BD/FFP formation mechanisms. 

Observationally, it is difficult to unambiguously identify young ejected objects, because simulations show that the ejection velocities are low, typically in the range 1--6~\kms\ \citep[e.g., Fig. 16 of][]{Stamatellos2009}. Adopting a sky projected velocity of $\sim3$~\kms, and for a protostellar lifetime of $\sim10^5$~yr, an ejected object might have travelled around 50000 au or about 2\arcmin\ at 400 pc, making its identification as an ejected object very challenging.
%say '1e5*3.1557e7*3e5/1.4960e13' = 6.3282754010695E+04
However, if the object was ejected within the last 10000 yr, it would be expected to be found within 12\arcsec\ of the parental protostar, and a first strategy would be to search for proto-BDs/FFPs in the surroundings of protostellar objects. 
In order to identify ejected objects in their earliest stages of formation, a promising technique would be to conduct SKAO+ngVLA+ALMA astrometric studies in nearby clouds.
%, as successfully demonstrated by Hern\'andez-Garnica et al. (2026). 
Thanks to the superb sensitivity of ALMA2040, these kind of astrometric studies will be performed routinely and with higher astrometric accuracy because of higher SNRs.

In addition, one of the predictions from the simulations of disk fragmentation + ejection by \cite{Stamatellos2009} is that a percentage of the ejected objects retain their disks and these disks present an excess of mass with respect to the expected \Mdisk\ vs \Mstar\ relation from, e.g., \cite{Andrews2013}. 
%and can be inferred from \cite{Stamatellos2009}, where the masses of ejected BDs are $\sim25$~\mj, and the disks of these BDs have masses in the range 1--5~\msun, corresponding to Mdisk-Mstar ratio in the range 0.04--0.2. 
The Ophiuchus BD disks studied by \cite{Testi2016_BDdisks} are consistent with the excess predicted by the simulations, as shown in Fig. 3 of \cite{Stamatellos2015}. ALMA2040 would allow us to build samples of confirmed ejected BDs/FFPs, and the comparison of disk properties of ejected objects vs objects formed in-situ will provide further clues to disentangle the role of the star-like and planet-like scenarios.

Finally, simulations show that the planet-like scenario 
%can be compact ($\lesssim1$~au) binaries with mass ratios near 1, 
and dynamically perturbed disks produce specific multiplicity properties that differ from the predictions of the star-like case \citep{Stamatellos2009,Thies2011_perturbeddiskcapture, Thies2015}. A simulation of a BD-BD binary with semimajor axis of 2 au and 25~\mj\ each component, eccentricity of 0.2 and inclination of 30$^\circ$, at 140 pc, observed in 13 epochs spanning 7 years with an astrometric accuracy of 1 mas (comparable to the accuracy that ALMA2040 will achieve with a beam of $\sim$10 mas) shows that about half of the orbit would be covered and that the orbital parameters could be properly recovered. These ALMA2040 astrometric studies would allow us to identify ejected or dynamically perturbed objects and study their multiplicity properties, providing valuable data to compare with the outcomes for each formation scenario \citep[e.g.,][]{Kroupa2003_OriginBDsFFPs, Marks2017_BDformation-binaries, Parker2023}.

\section{Planet formation around young BDs}\label{sec:planetform}

Although planets have been detected in a few cases around BDs \citep[e.g.,][]{Curiel2020_planet-BD}, and disks are expected to surround BDs and FFPs in the different formation scenarios,  
%The formation of planets around BDs requires a deep understanding of the physical processes on-going in the gaseous and dusty disk around these very low-mass objects \citep[e.g.,][]{Sanchez2022_planets-BDs}.
it is not clear how planet formation can succeed around very low-mass objects. This is because radial drift is supposed to be more efficient for lower-mass objects, promoting the quick drift of dust towards the inner disk. The presence of pressure bumps, very low disk gas masses or fast growth of porous and icy dust have been proposed to overcome this problem \citep[e.g.,][]{Pinilla2017}.

Alternatively, gravitational instability could favour planet formation further away from the inner disk. For this mechanism to be feasible, a \Mdisk/\Mstar\ ratio $\gtrsim0.3$ in the Class 0/I phase is required \citep[e.g.,][]{MercerStamatellos2020}. 
Although the typical observed ratio in Class II objects is $\sim0.001$ \citep[e.g.,][]{AlmendrosAbad2024_Maccrate-Mstar-substellar}, \Mdisk\ must have been 3--4 times larger in the previous Class 0/I phases \citep[e.g.,][]{Tychoniec2020, Tobin2020_VANDAM-Orion}, thus increasing the \Mdisk/\Mstar\ ratio. Therefore, at least for specific cases, the disk fragmentation scenario could be at work. 
%For example, Hernandez-Garnica et al. (2026) report a highly substructured disk around a proto-BD with an estimated \Mdisk/\Mstar\,$\sim0.5$. 
%\cite{MercerStamatellos2020} present simulations of gravitational instability around stars of a range of masses, being of 0.2~\ms\ stars the least massive star. 

\begin{figure*}[ht!]
\begin{center}
\includegraphics[width=0.32\columnwidth]{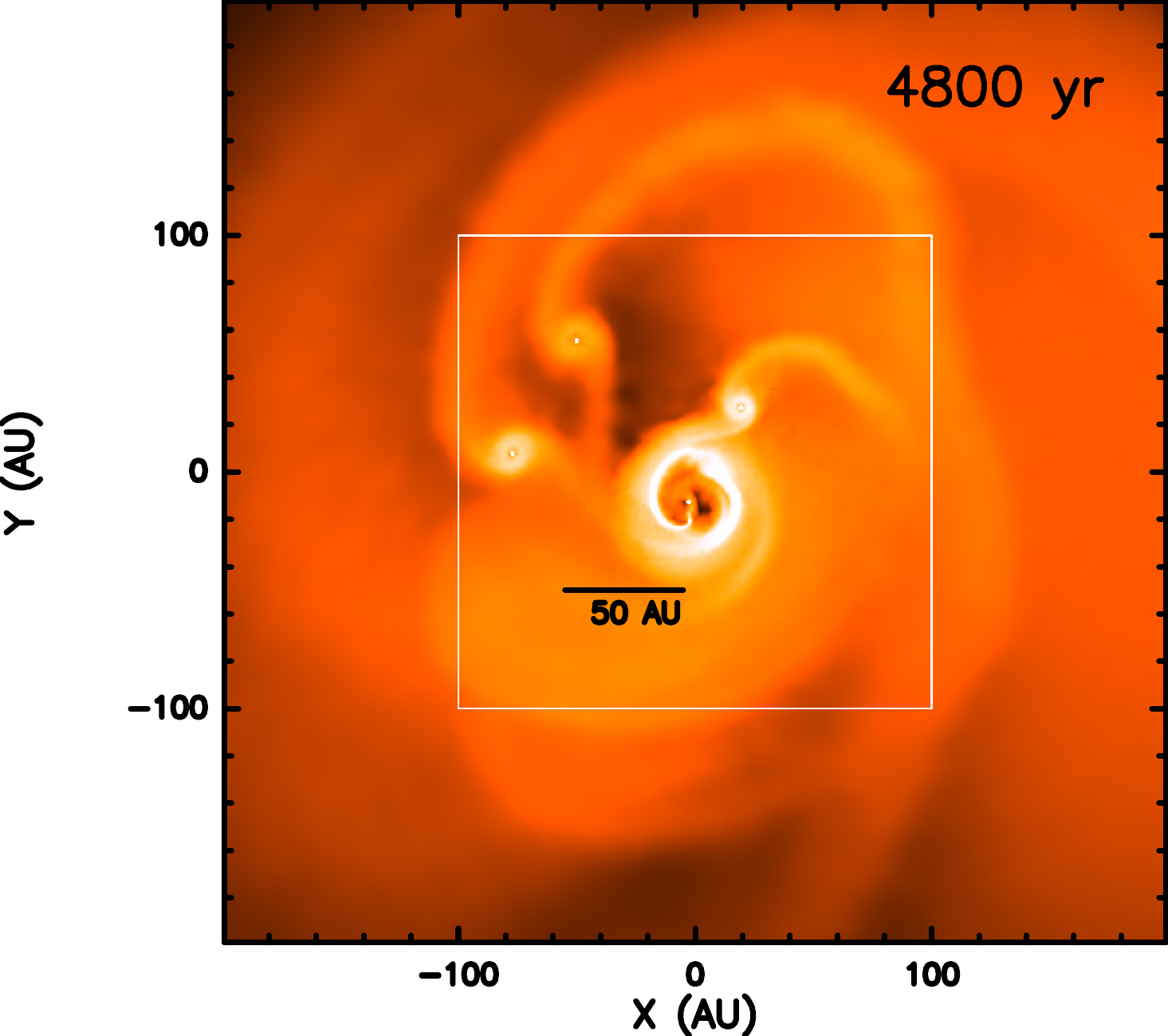}
\includegraphics[width=0.32\columnwidth]{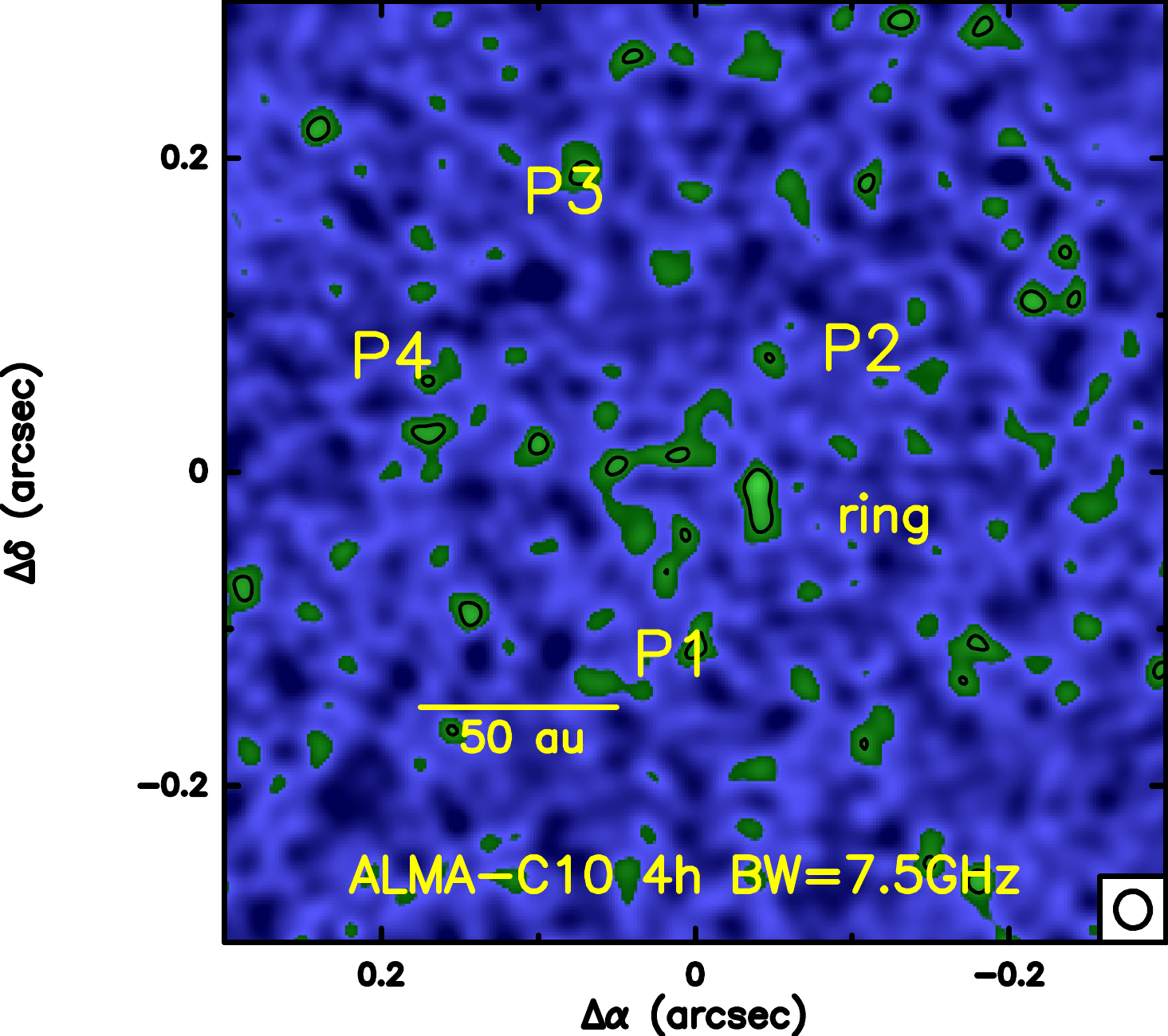}
\includegraphics[width=0.32\columnwidth]{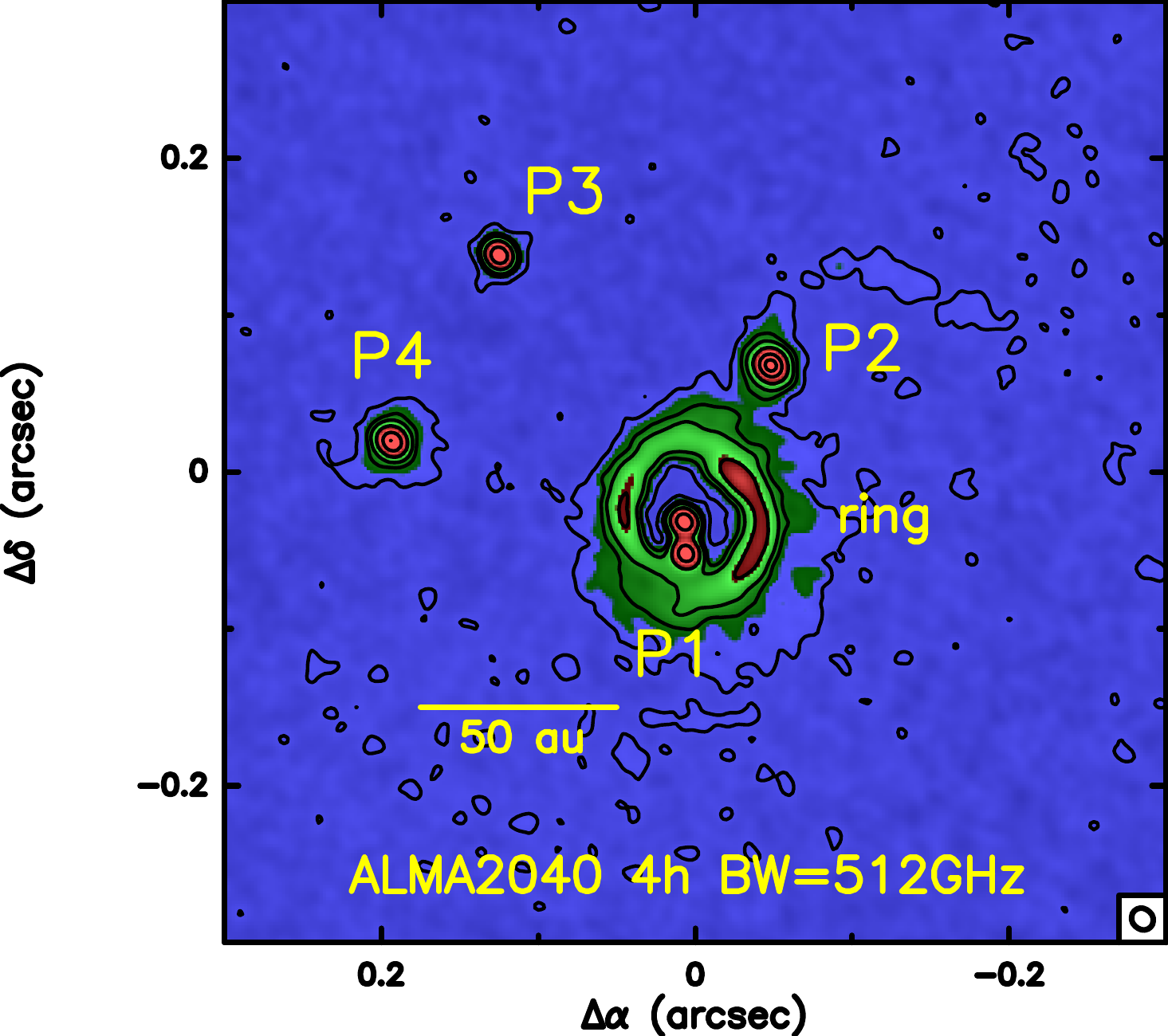}
\end{center}
\caption{{\bf Left:} Column density from Stamatellos et al. simulations at solar metallicity of the fragmenting disk around a 0.07~\ms\ star at 4800 yr. Substructure and planets are formed, with planets having their own disks. 
{\bf Center:} Zoom-in of the central white square of the simulation on left panel for 4 h of integration time with current ALMA.
%ALMA-C10: Restoring beam = 0.0236346" X 0.0218535", 7.70061 deg
%{\bf Bottom-left:} Zoom-in of the central white square of the simulation on the top-left panel as processed by GHOST to produce the synthetic observation. The ALMA2040 extended configuration was used for an integration time of 2 hours. The three outer planets, P2, P3 and P4  along with the innermost planet, P1, are detected above 7$\sigma$. In addition, the ring is detected at 5$\sigma$. Beam: $0.0145\times0.0126$ arcsec$^2$, 37.14 deg.
{\bf Right:} Idem for ALMA2040. 
%Beam: $0.015\times0.013$ arcsec$^2$, 44.5$^\circ$. 
}
\label{fig:Stamatellos} 
\end{figure*}

In order to explore the disk fragmentation scenario in a BD disk, a simulation of a gas fragmenting disk surrounding a 0.07~\msun\ object has been performed. The disk mass is 0.06~\msun\ and the disk radius is 80 au. Fig.~\ref{fig:Stamatellos}-left shows the column density from these solar-metallicity simulations for the 4800 yr snapshot. Substructure and fragments/planets are developed, with planets having their own disks. In particular, at this snapshot the simulation has developed four planets that are stable and persist during all the simulation. One of the planets lies at 8 au from the BD, and a ring of radius 15--17 au encompasses the BD+planet system. In addition, there are multiple spiral structures and three outer planets at 45--82 au from the BD, of masses 4--10~\mj. 
%(90, 165, 154 px)  

The simulated column density image was converted to intensity following the same assumptions of Sec.~\ref{sec:MCnumbers}, adopting a dust opacity per unit mass of dust+gas at 1.3 mm of 0.03 cm$^2$\,g$^{-1}$ \citep{Birnstiel2018_dustopacity}, and a distance of 400 pc.
Both ALMA in C10 configuration and ALMA2040 synthetic observations were performed for an integration time of 4 h and using 7.5 GHz of bandwidth for ALMA and 512 GHz for ALMA2040. %obtained using GHOST (\url{https://alma-ghost.jb.man.ac.uk}) 
The results are shown in Fig.~\ref{fig:Stamatellos}. The central panel of the figure shows that ALMA cannot detect any feature of this system, with an rms$=4$~$\mu$\jpb. On the contrary, ALMA2040, with rms$=0.12$~$\mu$\jpb, recovers the central source with its inner disk, its close planet P1, the circumbinary ring surrounding them, and the three forming planets further out (P2, P3, P4) at $>80\sigma$, along with their disks. 
%Planets P2, P3 and P4 have their own disks, with fluxes of 19.4, 11.8 and 14.5 $\mu$\jpb. This corresponds to circumplanetary masses of 0.13, 0.077, 0.095~\mj\ (for the same assumptions given above) for P2, P3 and P4, which are planets of 10, 8 and 4~\mj, respectively. 
Since ALMA2040 will allow us to detect planets of 0.002~\mj\ ($\sim 1$~\mearth, Table~\ref{tab:requirements}) and disk substructure around BDs/FFPs, planet demographics and planet formation pathways in the substellar regime will be constrained.

%\section{Multiplicity} 
%Multiplicity of BDs and free-floating planets \citep[e.g.,][]{Huelamo2026_IRAM04191, Bouy2026_Multiplicity-Taurus-UpSco} can provide decisive clues on their formation mechanism. ALMA2040 will allow to resolve and identify an important number of these systems. For example, it will allow to resolve compact binaries at distances XXX.
%Pennell, Rix et al. arXiv:2609.01822: Mass-dependent multiplicity fraction in low mass stars revealed by Gaia astrometry: the inferred multiplicity fraction is approximately constant at ∼45\% from the solar-type regime down to 0.4Msun, only then dropping to ∼15\% by 0.2Msun

%\section{Influence of the magnetic field on the formation and evolution} 
%The role of the magnetic field in the formation and chemistry of very low-mass objects

%\section{ALMA2040: Technical Requirements}
%As outlined in the previous sections, we summarize here the main technical requirements (angular resolution, spectral resolution, bandwidth, sensitivity) for each of the scientific questions presented above, given in Table~\ref{tab:requirements}. As can be seen from the table, ALMA2040 will allow to take a decisive step forward on each one of the science goals in feasible integration times.

%In order to detect pre-BD cores at distances one order-of-magnitude larger, of $\sim1$~kpc, a sensitivity at 3 mm of $\sim1$~$\mu$\jpb.
%In order to detect planets at 10$\sigma$ formed by disk fragmentation around BDs, at distances of $\sim400$~pc, a sensitivity at 1 mm of $\sim1.5$~$\mu$\jpb\ would be required.

\begin{table}
\small
	\centering
	\caption{Requirements for ALMA2040 to achieve the scientific goals of this white paper}
	\label{tab:requirements}
	\begin{tabular}{lccccl} % four columns, alignment for each
		\hline
        \hline
		Scientific goal & Distance$^\mathrm{a}$ &Sensitivity & Res.$^{\mathrm{b}}$ &BW$^{\mathrm{c}}$ &Int. \\
        to be achieved with ALMA2040 &(kpc) &($\mu$Jy/beam) & (mas) & (GHz) &time\\
		\hline
        Pre/proto-BD/FFP of $\sim1$~\mj\ numbers in MCs &$\lesssim10$ &0.1@3mm &20 &512 &1 h\\
        Idem covering nearby MCs with large mosaics &$\lesssim0.4$ &4@3mm &500 &512 &5s$^\mathrm{a}$\\
%        2. Detection of pre-BDs of $\sim20$~\mj &$\lesssim5$ &0.2 &40 &512 &1 h\\
        Detection of FH-BDs CO\,(2--1) outflows &$\lesssim0.14$ &100@1mm &500 &0.4$^{\mathrm{c}}$ &10 h\\
        Detection of $\sim1$~\mj\ FFP CO\,(2--1) outflows &$\lesssim0.14$ &100@1mm &500 &0.4$^{\mathrm{c}}$ & 10 h\\
        Detection of $\sim2$~\mearth\ disks of Class II/III BDs &$\lesssim1$ &0.2@1mm &5 &512 &1 h\\
        Resolving COMs in proto-BDs &$\lesssim0.4$ &200@3mm &10 &0.4$^{\mathrm{c}}$ &6 h\\
%        7. Detection of ejected proto-BDs &$\lesssim3$ & & \\
        Detection of $\lesssim1$~\mearth\ protoplanets around BDs &$\lesssim0.4$ &0.2@1mm &10 &512 &1 h\\
%        9. Detection of multiple proto-BD systems &$\lesssim3$ & & \\
		\hline
	\end{tabular}
    % For 7300, we changed the 3 sigma upper limit (<0.014) to 4sigma (0.152)
    \begin{flushleft}
    	$^{\mathrm{a}}$ Maximum distance where the science goal will be addressed with ALMA2040. If a single pointing is enough, ALMA2040 could conduct searches of BD+FFP up to $\sim10$~kpc. In order to conduct blind searches of BDs+FFPs in nearby clouds (at 140--400~pc), about 30000 pointings are required. \\
        $^{\mathrm{b}}$ To estimate the required angular resolution, a typical size of envelope/outflows of 200 au is adopted.\\
        $^{\mathrm{c}}$ Adopted bandwidth. For line observations, the number corresponds to the channel width in \kms.\\
        \end{flushleft}
\end{table}

%\subsection{Simulations with GHOST}
%Describe the configurations used to obtain the mock observations from GHOST at \url{https://alma-ghost.jb.man.ac.uk} and present the results.

\section{Synergies with current and future facilities}

%The ALMA2040 sensitivity will allow to detect radio emission from exoplanets and BDs as mentioned also for the case of the SKA by \cite{Kavanagh2026_SKA}.

ALMA2040 will be a crucial complement for a number of current and future facilities.
At radio wavelengths, next-generation interferometers such as the ngVLA, observing at wavelengths longer than 3~mm, will offer a unique opportunity to complement ALMA2040 searches in the northern hemisphere as their sensitivities at 100 GHz will be comparable. %The ngVLA sensitivity at 3 mm for 1 h integration with the Main + Long Baseline Arrays is 0.7~$\mu$\jpb, corresponding to a 5$\sigma$ mass sensitivity (adopting the same assumptions as in Sec.~\ref{sec:planetform}) of 1~\mj\ at 1 kpc, while ALMA2040 will be able to detect 0.2~\mj\ at 1 kpc at the same wavelength. Thus, samples with comparable sensitivities will be possible.
In addition, the ngVLA and SKAO will be able to conduct deep searches of radiojets driven by proto-BDs \citep[e.g.,][]{PerezGarcia2025_SUCANES}, and to search for gyrosynchrotron emission in the most evolved objects, allowing to conduct complementary studies to further constrain the nature of the objects.

%Phillipe André
%In the search of pre/proto-BDs/FFPs, wide-field multi-wavelength (0.35--1.3 mm) continuum surveys with large single-dish telescopes such as AtLAST, that can achieve beams of $\sim2$~arcsec, could be very efficient to identify candidate pre-BDs and proto-BDs in shorter amounts of times than ALMA2040. The most promising candidates could then be confirmed and further characterized with ALMA2040.

On the mid-infrared side, the JWST will be an excellent complement to further constrain the nature of the central objects and the composition of their surrounding disks. For example, \cite{Cugno2025_JWST-C-O-ratios} report a carbon-rich disk surrounding a planetary-mass object.
%Carlos Carrasco-Gonzalez
In addition, recent studies suggest that lower-mass objects convert accretion into ejection more efficiently \citep[e.g.,][]{Rota2025_CorrelationMaccMout}, hinting at mass-dependent accretion mechanisms or distinct evolutionary pathways in the substellar regime. To test this, combining ALMA2040 dust continuum observations with ngVLA free-free emission from winds/jets and JWST accretion tracers will enable a comprehensive multi-wavelength characterization of the accretion-ejection connection in BDs. 
%Extending this approach to the lowest mass objects will place crucial observational constraints on how mass accumulation and outflow launching scale down across the substellar boundary.

Overall, the synergy between ALMA2040, ngVLA, SKAO, and JWST 
%and large single-dishes such as AtLAST 
will constitute a  powerful combination to elucidate how common BDs and FFPs are in our Galaxy, which main formation mechanisms have produced them, what type and how many planets do we expect around them and what chances we have to detect pre-biotic chemistry in their disks.

\section*{Acknowledgements}
This work was developed as part of the broader ALMA2040 community effort. We acknowledge the Lorentz Center and NOVA for supporting the community workshops and coordination activities that helped initiate and shape the ALMA2040 White Paper process.
A.P. acknowledges financial support from the UNAM-PAPIIT IN120226 grant, M\'exico.
\\

\bibliography{browndwarfs_refs}
\bibliographystyle{apj_mod}

\newpage

\section*{Affiliations}

\vspace{1em}
{
\footnotesize{
\noindent $^1$ Instituto de Radioastronomía y Astrofísica, Universidad Nacional Autónoma de México, México: \Letter \url{a.palau@irya.unam.mx}; 
$^2$ Jeremiah Horrocks Institute for Mathematics, Physics and Astronomy, University of Lancashire, UK; 
$^3$ Department of Physics and Astronomy, University of Exeter, UK; 
$^4$ Instituto de Astrofísica, Óptica y Eletrónica, México; 
$^5$ Centro de Astrobiología (CAB), CSIC-INTA, Madrid, Spain; 
$^6$ School of Physical and Chemical Sciences, Queen Mary University of London, UK;
$^{7}$ Laboratoire d’Astrophysique (AIM) Université Paris-Saclay, Université Paris Cité, CNRS, France;
$^{8}$ Instituto de Astrofísica de Canarias, Spain;
$^{9}$ Max-Planck Institute for Astronomy (MPIA), Heidelberg, Germany;
$^{10}$ Institut de Recherche en Astrophysique et Planétologie, Université de Toulouse, CNRS, Toulouse, France;
$^{11}$ National Astronomical Observatory of Japan, National Institutes of Natural Sciences, Osawa, Mitaka, Tokyo Japan; 
$^{12}$ Helmholtz Institute for Radiation and Nuclear Physics, University of Bonn, Germany;
$^{13}$ SUPA, School of Physics \& Astronomy, St Andrews, UK;
$^{14}$ Astrophysics Research Cluster, School of Mathematical and Physical Sciences, The University of Sheffield, UK;
$^{15}$ European Southern Observatory, Garching, Germany;
$^{16}$ Observatorio Astronómico Nacional, Spain;
$^{17}$ ATNF CSIRO Space and Astronomy, Australia;
$^{18}$ School of Physics, Trinity College Dublin, the University of Dublin, Ireland;
$^{19}$ Astronomical Institute, Tohoku University, Japan; 
$^{20}$ Escuela Nacional de Estudios Superiores (ENES) Unidad Morelia, Universidad Nacional Autónoma de México, Morelia, México;
$^{21}$ Institut de Ciències de l’Espai (ICE-CSIC), Spain;
$^{22}$ European Southern Observatory, Santiago, Chile;
$^{23}$ Dipartimento di Fisica “Aldo Pontremoli”, Università degli Studi di Milano, Milano, Italy;
$^{24}$ National Astronomical Observatory of Japan, Mitaka, Tokyo, Japan;
$^{25}$ Leiden Observatory, Leiden University, The Netherlands;
$^{26}$ Department of Astrophysics, American Museum of Natural History, New York, USA;
$^{27}$ Department of Physics, International University, Ho Chi Minh City, Vietnam;
$^{28}$ Univ Lyon, Centre de Recherche Astrophysique de Lyon, France;
$^{29}$ SUPA, School of Science and Engineering, University of Dundee, Dundee, UK;
$^{30}$ Departamento de Fisica Aplicada III, ETSI, Sevilla University, Spain;
$^{31}$ University College London, UK; 
$^{32}$ LIRA, Observatoire de Paris, Université PSL, Sorbonne Université, Université Paris Cité, CNRS, Meudon, France;
$^{33}$ Division of Geological and Planetary Sciences, California Institute of Technology, Pasadena, USA; 
$^{34}$ Department of Astronomy, University of Texas at Austin, Austin, TX, USA;
$^{35}$ Maryland Center for Fundamental Physics, University of Maryland, College Park, USA;
$^{36}$ School of Physics and Astronomy, University of Leeds, UK;
$^{37}$ Laboratoire d’Astrophysique de Bordeaux, Université de Bordeaux;
$^{38}$ Department of Astronomy \& Astrophysics, UC San Diego, USA;
$^{39}$ Department of Astronomy, Boston University, USA;
$^{40}$ Institute of Astronomy, Department of Physics, National Tsing Hua University, Hsinchu, Taiwan;
$^{41}$ Department of Earth and Planetary Sciences, Faculty of Sciences, Kyushu University, Japan;
$^{42}$ Instituto de Astrofísica e Ciências do Espaço, Faculdade de Ciências, Universidade de Lisboa, Lisbon, Portugal;
$^{43}$ I. Physikalisches Institut, Universität zu Köln, Germany;
$^{44}$ Instituto de Astronom\'ia, Universidad Nacional Aut\'onoma de M\'exico;
$^{45}$ Homer L. Dodge Department of Physics and Astronomy, University of Oklahoma,  USA;
$^{46}$ CHART, School of Physics and Astronomy, Cardiff University, Cardiff, UK.
}
}

\end{document}